\documentclass[%
 superscriptaddress,
showkeywordspreprint,
tightenlines,
showpacs,preprintnumbers,
amsmath,
amssymb,
graphicx,bbm,subcaption,xcolor,braket,latexsym,microtype
 longbibliography,
pra,
]{revtex4-1}

\usepackage{graphicx}
\usepackage{dcolumn}
\usepackage{amsmath,amsfonts,amsthm,bm}
\usepackage{bbm}
\usepackage{subfigure}

\usepackage{color}
\usepackage[resetlabels,labeled]{multibib}

\begin{document}


\title{ON THE QUANTUMNESS OF MULTIPARAMETER ESTIMATION PROBLEMS FOR QUBIT SYSTEMS}

\author{Sholeh Razavian}
\email{sholeh.razavian@gmail.com}
\affiliation{%
Faculty of Physics, Azarbaijan Shahid Madani University, Tabriz 5375171379, Iran
}%
\affiliation{
Quantum Technology Lab, Dipartimento di Fisica ``Aldo Pontremoli'', Universit\`a degli Studi di Milano, I-20133 Milano, Italy
}%
\author{Matteo G. A. Paris}%
\email{matteo.paris@fisica.unimi.it}
\affiliation{%
Quantum Technology Lab, Dipartimento di Fisica ``Aldo Pontremoli'', Universit\`a degli Studi di Milano, I-20133 Milano, Italy
}%
\affiliation{%
INFN, Sezione di Milano, I-20133 Milano, Italy
}%

\author{Marco G. Genoni}
\email{marco.genoni@fisica.unimi.it}
\affiliation{
Quantum Technology Lab, Dipartimento di Fisica ``Aldo Pontremoli'', Universit\`a degli Studi di Milano, I-20133 Milano, Italy
}%
\affiliation{
INFN, Sezione di Milano, I-20133 Milano, Italy
}%
\date{\today}

\begin{abstract}
The estimation of more than one parameter in quantum mechanics is a fundamental problem with relevant practical applications. In fact, the ultimate limits in the achievable estimation precision are ultimately linked with the non-commutativity of different observables, a peculiar property of quantum mechanics. We here consider several estimation problems for qubit systems and evaluate the corresponding {\em quantumness} $\mathcal{R}$, a measure that has been recently introduced in order to quantify how much incompatible are the parameters to be estimated. In particular, $\mathcal{R}$ is an upper bound for the renormalized difference between the (asymptotically achievable) Holevo bound and the SLD Cram\'er-Rao bound (i.e. the matrix generalization of the single-parameter quantum Cram\'er-Rao bound). For all the estimation problems considered, we evaluate the quantumness $\mathcal{R}$ and, in order to better understand its usefulness in characterizing a multiparameter quantum statistical model, we compare  it with the renormalized difference between the Holevo and the SLD-bound. 
Our results give evidence that $\mathcal{R}$ is a useful quantity to characterize multiparameter estimation problems, as for several quantum statistical model it is equal to the difference between the bounds and, in general, their behaviour qualitatively 
coincide. On the other hand, we also find evidence that for certain quantum statistical models the bound is not in tight, and thus $\mathcal{R}$ may overestimate the degree of quantum incompatibility between parameters.
\end{abstract}
\maketitle
\section*{Introduction}
Quantum sensing is the art of exploiting quantum features as coherence (or decoherence) 
to improve the sensitivity of measuring devices \cite{Giovannetti1330,Paris2009,Giovannetti2011,Demkowicz-Dobrzanski2015a,Degen2016,Braun2018,Pezze2018,PhysRevD.94.024014,PhysRevA.93.053805,Rossi_2017,PhysRevLett.124.120504}. The field originates from 
fundamental research in the detection of gravitational waves, and is now a successful
quantum technology. When the problem at hand involves a single parameter, there is
a clear avenue to define optimality: build a suitable quantum statistical model, evaluate the 
symmetric logarithmic derivative (SLD) operator and then the corresponding
quantum Cram\'er-Rao SLD-bound.
The SLD-bound indeed sets the ultimate
precision achievable by any estimation strategy, and links it 
to the quantum Fisher Information (QFI), which quantifies the amount of information extractable via  
quantum measurements on the parameter. 
Quantum probes and detectors are then designed to maximize and possibly attain 
the QFI for given resources, possibly outperforming any conceivable 
setup using only classical states and detectors.
\par
As a matter of fact, there are several problems of interest that are inherently involving more than one parameter 
\cite{AlbarelliPerspective,RafalReview2020,Liu_2019,Szczykulska2016}, e.g. estimation of unitary operations and of multiple phases \cite{Ballester2004,Vaneph2012,Genoni2013b,Humphreys2013,de_Falco_2013,Tamascelli_2014,Gagatsos2016a,Knott2016,Pezze2017,Roccia2017,friel2020}, estimation of phase and noise \cite{Vidrighin2014,Crowley2014,Altorio2015a,Roccia2018}, and superresolution of incoherent sources~\cite{Chrostowski2017,Yu2018,Napoli2018}. However, 
despite these important applications, multiparameter quantum estimation received less 
attention and some relevant (fundamental and practical) issues have not yet fully resolved. In particular, 
the non-commutativity of the quantum observables needed to jointly estimate 
a vector of parameters poses non trivial limitations. As a result, despite 
its classical counterpart is asymptotically achievable, non-commutativity \cite{Zhu,Heinosaari2016}, makes the multiparameter SLD-bound \cite{Helstrom1967,helstrom1976quantum} generally not achievable. Indeed, 
the multiparameter estimation bounds are given by matrix inequalities which 
are not tight. In order to access the fundamental precision limit 
in multiparameter quantum estimation problem, Holevo proposed a scalar bound \cite{Holevo1973,Holevo2011b}.
The Holevo bound is regarded as the most fundamental scalar lower bound, as it is attainable by allowing collective measurements on an asymptotically large number of copies of the quantum state defining the quantum statistical model~\cite{Hayashi2008a,Guta2006}. 
Besides the practical applications, the difference between the Holevo- and the  SLD-bound also allows to assess 
the degree of {\em incompatibility} of the parameters to be estimated. Nevertheless, the expression of the Holevo bound 
is not crystal clear in terms of the model under consideration, since it is 
written as an optimization over a set of matrices.
\par
We take the simplest quantum systems, a qubit, and consider several multiparameter
estimation problems involving unitary and noisy channels. For those models,
we evaluate its {\em quantumness} $\mathcal{R}$, that has been recently 
introduced in \cite{Carollo2019}. This quantity is supposed to estimate the amount of the incompatibility
of the parameters defining the quantum statistical model and can be easily evaluated via the Uhlmann 
curvature matrix and the SLD-QFI matrix. Then we also evaluate the Holevo bound that, for the case of two-parameter qubit models, can be evaluated analytically via the SLD and the
right logarithmic derivative (RLD) operators only \cite{Suzuki2016a}.
In particular, we compare $\mathcal{R}$ with the renormalized difference $\Delta C_{\sf max}$ between 
the Holevo and the SLD-bound, maximized over all the possible weight matrices, 
to assess its performance as an upper bound.
Our results show that the two quantities share always the same qualitative behaviour, but also give evidence that
the bound is not always tight, i.e. $\mathcal{R}$ 
is a useful quantity to characterize multiparameter estimation problems, 
but it may overestimate the incompatibility between parameters.
\par
The paper is structured as follows. In Sec. 2, we briefly review the theory behind multi-parameter quantum metrology, we introduce the measure for {\em quantumness} $\mathcal{R}$ and discuss its main properties. In Sec. 3, we evaluate the quantumness parameter and the renormalized difference between Holevo- and SLD-bound for several multiparameter quantum statistical models for single qubits, discussing their relationship in more detail. Sec. 4 concludes the manuscript with a final discussion. 

 \section*{Multi-parameter quantum metrology and a measure of {\em quantumness} for quantum statistical models}
 
 In this section, we will provide all the basic notions of multi-parameter quantum metrology that are needed for our goals. We refer to the following references \cite{AlbarelliPerspective,RafalReview2020,Liu_2019,Szczykulska2016} for more explanations and technical details on this topic. \\
 The aim of local estimation theory is to set the ultimate bounds on how precisely the considered parameters can be estimated. For this purpose, general bounds have been proposed \cite{Helstrom1967,Yuen1973,Belavkin1976} which depend on the quantum statistical model
  $\varrho_{\bm{\lambda}}$ only, that is on the family of quantum states defined in terms of the set of parameters 
 ${\bm{\lambda}} = (\lambda_1,...\lambda_n)$ to be estimated. 
 In particular one introduces, respectively, the SLD operators $L^S_\mu$, and the
 RLD operators $L^R_\mu$ via the equations
 
 \begin{align}
 \partial_\mu\varrho_{\bm{\lambda}} &=\frac{L_\mu^S\,\varrho_{\bm{\lambda}} + \varrho_{\bm{\lambda}}\,L_\mu^S}{2} \,, \label{eq:SLD} \\
 \partial_\mu\varrho_{\bm{\lambda}}&=\varrho_{\bm{\lambda}}\,L_\mu^R \,,
 \label{eq:RLD}
 \end{align}

where $\partial_\mu$ corresponds to the partial derivative respect to the $\mu$-th parameter $\lambda_\mu$.
It is then possible to derive the following (measurement independent) matrix quantum Cram\'er-Rao bounds (CRB), bounding the covariance matrix $\bm V(\lambda)$ of any locally unbiased estimator \cite{Holevo2011b,Hayashi2008a,Gill2000}
\begin{equation}
{\bm V}({\bm{\lambda}})\geq {\bm Q}(\bm{\lambda})^{-1},\qquad\qquad {\bm V}({\bm{\lambda}})\geq {\bm J}({\bm{\lambda}})^{-1}
\end{equation}
where the corresponding SLD $\bf{Q(\bm{\lambda})}$ and RLD $\bf{J(\bm{\lambda})}$ quantum Fisher information (QFI) matrices elements are
\begin{align}
Q_{\mu\nu}({\bm{\lambda}})&=\text{Tr}\left [\varrho_{\bm{\lambda}}\frac{L_\mu^S\,L_\nu^S+L_\nu^S\,L_\mu^S}{2} \right] \,,
\label{SLDQFI}
\\
J_{\mu\nu}({\bm{\lambda}})&=\text{Tr}[\varrho_{\bm{\lambda}}\,L_\nu^R\,L_\mu^{R\dagger}] \,.
\label{RLDQFI}
\end{align}

In the single-parameter case, the SLD-bound is tight, that is it is always possible to saturate it by performing the optimal measurement, that in turn corresponds to projecting on the eigenstates of the SLD operator $L^S_\lambda$. 

In the multi-parameter scenario, it is common to rewrite the matrix bounds into scalar CRBs by introducing a positive, real {\em weight matrix} ${\bm W}$, leading to the inequalities
\begin{align}
\text{Tr}[{\bm{W\, V}}]\geq C^{\sf S}({\bm{\lambda}},{\bm W}),\qquad \text{Tr}[{\bm{W\, V}}]\geq C^{\sf R}(\bm{\lambda},\bm{W}) \,,
\end{align}
where the scalar SLD- and RLD-CRBs respectively read
\begin{align}
C^{\sf S}({\bm{\lambda}},{\bm W}) &=\text{Tr}[{\bm W} \bm{Q}^{-1}] \,, \\
C^{\sf R}({\bm{\lambda}},{\bm W}) &=\text{Tr}[{\bm W} {\rm Re}({\bm J}^{-1})]+||\sqrt{{\bm W}}{\rm Im}({\bm J}^{-1})\sqrt{{\bm W}}||_1
\end{align}
with $||{\bm A}||_1=\text{Tr}[\sqrt{{\bm A}^\dagger {\bm A}}]$ denoting the trace norm of the matrix ${\bm A}$. 
Holevo derived a tighter scalar bound $C^{\sf H}({\bm{\lambda}},{\mathbf W})$ in \cite{Holevo1973,Holevo2011b} via the following minimization
\begin{align}
C^{\sf H}(\bm{\lambda}, {\bf W}) &=  \min_{{\bf U} \in \mathbbm{S}^d, {\bf X} \in \mathbbm{X}_{\bm{\lambda}} }  \left[ \text{Tr}[ {\bf W}\,{\bf U} ] \,\, | \,\,{\bf U} \geq {\bf Z}[{\bf X}] \right] \\
&= \min_{{\bf X} \in \mathbbm{X}_{\bm{\lambda}} }  \left[ \text{Tr}[ {\bf W} \,  \text{Re} {\bf Z}[{\bf X}] ] + \Vert \sqrt{\mathbf{ W}} \,  \text{Im} {\bf Z}[{\bf X}] \sqrt{\mathbf{W}} \Vert_1  \right] \,, \label{eq:HolevoBound} 
\end{align}
where $\mathbbm{S}^d$ denotes the set of real symmetric $d$-dimensional matrices, and the Hermitian $d \times d$ matrix ${\bf Z}$ is defined via its elements
\begin{align}
Z_{\mu \nu} \left[{\bf X}\right]= \text{Tr}[ \varrho_{\bm{\lambda}} X_\mu X_\nu ] \,,
\end{align}
with the collection of operators ${\bf X}$ belonging to the set
\begin{align}
\mathbbm{X}_{\bm{\lambda}} = \left\{ {\bf X} = (X_1, \dots , X_d )\,\, |\,\,  \text{Tr}[( \partial_\mu \varrho_{\bm{\lambda}} ) X_\nu ]= \delta_{\mu\nu} \right\} \,.
\end{align}
In fact, the following chain of inequalities holds
\begin{align}
\text{Tr}[{\bf W\, V}]  \geq C^\mathsf{H}(\bm{\lambda}, {\bf W}) 
\geq \textrm{max}\left[ C^{\sf S}(\bm{\lambda}, {\bf W}), C^{\sf R}(\bm{\lambda}, {\bf W})\right] \,. \label{eq:chain}
\end{align}

The Holevo bound is achievable if one considers the corresponding asymptotic model, that is by optimizing over collective measurements on an asymptotically large number of copies of the state, $\varrho_{\bm{\lambda}}^{\otimes n} = \bigotimes_{j=1}^n \varrho_{\bm{\lambda}}$, with $n \to \infty$~\cite{Hayashi2008a,Kahn2009,Yamagata2013,Yang2018a}. It has been proved that there are instances where the Holevo bound is also achieved in the single-copy scenario: this is the case for example of pure state models~\cite{Matsumoto2002} and displacement estimation with Gaussian states~\cite{Holevo2011b}. 

The SLD- and the Holevo-bound stand out as the most exploited tools in order to characterize a multi-parameter estimation problem. On the one hand, $C^{\sf S} (\bm{\lambda} , {\bf W} )$ is the straightforward generalization of the single-parameter bound and it is typically easy to calculate. On the other hand $C^{\sf H} (\bm{\lambda} , {\bf W} )$, despite being difficult to calculate because of the complicated minimization procedure, is always more informative than $C^{\sf S} (\bm{\lambda} , {\bf W} )$ and it is achievable, at least for the asymptotic model. More recently it has been shown that the Holevo bound can be also upper bounded in terms of the SLD-bound, as follows \cite{Carollo2019,albarelli2019upper,Tsang2020}:
\begin{align}
C^{\sf S} (\bm{\lambda} , {\bf W} ) &\leq C^{\sf H}(\bm{\lambda}  , {\bf W}) \\
&\leq C^{\mathsf{S}}(\bm{\lambda} , {\bf W} ) + \Vert \sqrt{\mathbf{W}}  \, \mathbf{Q}^{-1} \, \mathbf{D} \, \mathbf{Q}^{-1}\, \sqrt{\mathbf{W}} \Vert_1  \\
&\leq ( 1 + \mathcal{R})\, C^{\sf S} (\bm{\lambda} , {\bf W} )  \leq 2 \, C^{\sf S} (\bm{\lambda} , {\bf W} )\,.
\label{eq:carollo}
\end{align}
In the chain of inequalities above we have introduced two new objects: the (asymptotic) \emph{incompatibility} matrix ${\bf D}$, also known as mean Uhlmann curvature~\cite{Carollo2018a},
with elements
\begin{align}
D_{\mu\nu} = -\frac{i}{2} \text{Tr}[ \varrho_{\bm{\lambda}} [L_\mu^{\sf S}, L_\nu^{\sf S}] ] \, ,
\end{align}
and the {\em quantumness} measure
\begin{align}
\mathcal{R} = \parallel i \, \mathbf{Q}^{-1} {\bf D} \parallel_\infty \, , \label{eq:Rdefinition}
\end{align}
where $\parallel {\bf A} \parallel_\infty$ denotes the largest eigenvalue of the matrix ${\bf A}$.
\subsection{On the quantumness parameter $\mathcal{R}$}
The figure of merit $\mathcal{R}$ introduced in \cite{Carollo2019} via the eq. (\ref{eq:Rdefinition}) is the main focus of our work. As we will describe below by revising and deriving some of its fundamental properties, $\mathcal{R}$ quantifies the {\em quantumness} of a multi-parameter quantum statistical model, or more in detail, the (asymptotical) incompatibility of the parameters to be estimated. 

As it is clear from its definition, $\mathcal{R}$ is well defined only for quantum statistical models $\varrho_{\bm{\lambda}}$ having a non-singular SLD-QFI matrix, and we will thus avoid these pathological cases. Here below we present a list of its properties (notice that we will provide the proof only for property [P5], as it was not presented in \cite{Carollo2019}):
\begin{itemize}
	\item[{\bf [P1]}]{the quantumness measure $\mathcal{R}$ is bounded as follows
		\begin{align}
		0 \leq \mathcal{R} \leq 1\,.
		\end{align}
	}
	\item[{\bf [P2]}] one has that
	\begin{align}
	\mathcal{R}=0 \,\, \Leftrightarrow {\bf D}=\bm{0} \,,
	\end{align} 
	that is if and only if the {\em weak compatibility} condition for the SLD operators holds, 
	$$\text{Tr}[\varrho_{\bm{\lambda}} [L_\mu^{\sf S}, L_\nu^{\sf S}]]=0\,.$$ 
	Consequently, in this case, one has that $C^{\sf H}(\bm{\lambda} , {\bf W} )=C^{\sf S}(\bm{\lambda} , {\bf W} )$ for all weight matrices ${\bf W}$, and thus the quantum statistical model is said to be {\em asympotically classical}: the SLD-bound is asymptotically achievable via collective measurements on $\varrho_{\bm{\lambda}}^{\otimes n} = \bigotimes_{j=1}^n \varrho_{\bm{\lambda}}$, with $n \to \infty$~\cite{Ragy2016}.
	\item[{\bf [P3]}] Given any possible weight matrix ${\bf W}$, the following inequality holds:
	\begin{align}
	\Delta C (\bm{\lambda}, {\bf W}) \leq \mathcal{R} \,,
	\end{align}
	that is the quantumness $\mathcal{R}$ is an upper bound for the renormalized difference between Holevo and SLD-bound
	\begin{align}
	\Delta C(\bm{\lambda} ,{\bf W}) = \frac{C^{\sf H}(\bm{\lambda} , {\bf W} )-C^{\sf S}(\bm{\lambda} , {\bf W} )}{C^{\sf S}(\bm{\lambda} , {\bf W} )} \,.
	\end{align}
	\item[{\bf [P4]}] If the number of parameters to be estimated is $n=2$, one has that
	\begin{align}
	\mathcal{R} =\sqrt\frac{\det {\bf D}}{\det {\bf Q}} \,. \label{eq:R2parameter} 
	\end{align}
	\item[{\bf [P5]}]
	The quantumness $\mathcal{R}$ is invariant under reparametrization of the quantum statistical model: given a new statistical model
	$\varrho_{\bar{\bm{\lambda}}}$, such that the new set of $n$ parameters are obtained as a function of the original ones,
	$\bar{\bm{\lambda}}=f(\bm{\lambda})$, then
	\begin{align}
	\mathcal{R}(\bar{\bm{\lambda}})=\mathcal{R}(\bm{\lambda})\,.
	\end{align}
	{\em Proof} - The SLD- and Uhlmann curvatures matrices for the two quantum statistical models are related via the equations
	\begin{align}
	\mathbf{Q}(\bar{\bm{\lambda}}) = {\bf B} \,\mathbf{Q}(\bm{\lambda}) \,{\bf B}^{\sf T},\,\,\,\,\, {\bf D}(\bar{\bm{\lambda}}) = {\bf B} \,{\bf D}(\bm{\lambda}) \,{\bf B}^{\sf T}  \,,
	\end{align}
	where the reparametrization matrix ${\bf B}$ is defined via its elements $B_{\mu\nu} = \partial \lambda_\nu / \partial\bar{\lambda}_\mu$. As a consequence one can write the corresponding quantumness parameter as 
	\begin{align}
	\mathcal{R}(\bar{\bm{\lambda}}) &= \parallel i \mathbf{Q}(\bar{\bm{\lambda}})^{-1} \mathbf{D}(\bar{\bm{\lambda}})\parallel_{\infty}\\
	&=\parallel i (\mathbf{B}^{T})^(-1) \mathbf{Q}({\bm{\lambda}})^{-1} \mathbf{B}^{-1}\mathbf{B} \mathbf{D}({\bm{\lambda}})\mathbf{B}^T \parallel_{\infty}\\
	&=\parallel i (\mathbf{B})^{-1} \mathbf{Q}({\bm{\lambda}})^{-1} \mathbf{D}({\bm{\lambda}})\mathbf{B}^{T}\parallel_{\infty}\\
	&= \parallel i \mathbf{Q}({\bm{\lambda}})^{-1} \mathbf{D}({\bm{\lambda}})\parallel_{\infty}\\
	&=\mathbf{R}({\bm{\lambda}})\,,
	\end{align}
	where the next-to-last equality is satisfied as {\em similar matrices} (a matrix $A$ is similar to a matrix $B$ if an invertible matrix $C$ exists such that $B=C^{-1} A C$) have the same eigenvalues. 
	One should notice that this result is in fact consistent with the observation that the quantity $\mathcal{R}$ does not depend on the weight matrix ${\bf W}$: to apply a different weight matrix is formally equivalent to define a new quantum statistical model via a reparametrization of the set of parameters $\bm{\lambda}$.
\end{itemize}
\subsection{On the evaluation of the Holevo bound for single qubit statistical models}
In the next section, we will evaluate the quantumness parameter $\mathcal{R}$ and the renormalized difference between Holevo- and SLD-bound $\Delta C(\bm{\lambda}, {\bf W})$ for several multiparameter quantum statistical models for single qubits.  In order to calculate $\Delta C(\bm{\lambda}, {\bf W})$ it will be necessary to evaluate both the SLD-bound and the Holevo bound. While the evaluation of the SLD-bound is typically straightforward, as we mentioned before the minimization needed for the evaluation of the Holevo bound is in general complicated. There are however few instances where the Holevo bound can be always easily calculated: 
\begin{itemize}
	\item {\em Asymptotically classical models}: as previously discussed if ${\bf D}(\bm{\lambda})=\bm{0}$, then one has straightforwardly that $C^{\sf H}(\bm{\lambda} , {\bf W} )=C^{\sf S}(\bm{\lambda} , {\bf W} )$.
	\item {\em D-invariant models}: if a model is D-invariant (we refer to these references \cite{Suzuki2018,AlbarelliPerspective} for a precise definition and characterization of quantum statistical models, as it goes beyond the scope of this work), then 
	\begin{align}
	C^{\sf H}(\bm{\lambda} , {\bf W} ) &= C^{\sf R}(\bm{\lambda} , {\bf W} ) \\
	&=C^{\sf S}(\bm{\lambda} , {\bf W} ) +  \Vert \sqrt{\mathbf{W}}  \, \mathbf{Q}^{-1} \, \mathbf{D} \, \mathbf{Q}^{-1}\, \sqrt{\mathbf{W}} \Vert_1 \,, \label{eq:HolevoDInvariant}
	\end{align}
	that is the Holevo-bound is equal to the RLD-bound and both can be expressed in terms of the SLD matrices ${\bf Q}$ and ${\bf D}$ only \cite{Suzuki2016a}. It is important to remark that all quantum statistical models corresponding to full state tomography of finite-dimensional quantum system are D-invariant.
\end{itemize}
If the quantum statistical model does not fall into these classes, the evaluation of $C^{\sf H}(\bm{\lambda} , {\bf W} )$ may be unfeasible. Several efforts have been made in the literature in order to obtain numerical or even analytical results, at least for some specific classes of quantum states \cite{Matsumoto1997,Crowley2014,Baumgratz2015,Bradshaw2017a,Bradshaw2017,Sidhu2018,Albarelli2019,friel2020}. In particular, a closed formula has been derived for all single-qubit two-parameter quantum statistical models \cite{Suzuki2016a}. In order to obtain it, we have first to introduce a new quantity, namely
\begin{align}
C^{\sf Z}({\bm{\lambda}},{\mathbf W}) = C^{\sf S}(\bm{\lambda} , {\bf W} ) +  \Vert \sqrt{\mathbf{W}}  \, \mathbf{Q}^{-1} \, \mathbf{D} \, \mathbf{Q}^{-1}\, \sqrt{\mathbf{W}} \Vert_1 \,.
\end{align}
The Holevo bound for any two-parameter qubit model under the regularity condition can then be written as
\begin{equation}
C^{\sf H}({\bm{\lambda}},{\mathbf W})=\left \{ \begin{array}{cc}
C^{\sf R}({\bm{\lambda}},{\mathbf W})& \mbox{if}\,\, C^{\sf R}({\bm{\lambda}},{\mathbf W})\geq\frac{C^{\sf Z}({\bm{\lambda}},{\mathbf W})+C^{\sf S}({\bm{\lambda}},{\mathbf W})}{2}\\
C^{\sf R}({\bm{\lambda}},{\mathbf W})+S({\bm{\lambda}},{\mathbf W})& \mbox{otherwise}
\label{eq:Holevoqubit}
\end{array}
\right.
\end{equation}
where the function $S({\bm{\lambda}},{\mathbf W})$ is non-negative and defined by
\begin{equation}
S({\bm{\lambda}},{\mathbf W}):=\frac{[\frac{1}{2}(C^{\sf Z}({\bm{\lambda}},{\mathbf W})+C^{\sf S}({\bm{\lambda}},{\mathbf W}))-C^{\sf R}({\bm{\lambda}},{\mathbf W})]^2}{C^{\sf Z}({\bm{\lambda}},{\mathbf W})-C^{\sf R}({\bm{\lambda}},{\mathbf W})}.
\end{equation}
From the formulas above, it is apparent that the Holevo bound for two-parameter estimation problems can be written in terms of SLD and RLD operators only.
\section{{\em Quantumness} of single-qubit multiparameter quantum statistical models}
In this section, we will present the main results of our manuscript. We will consider different quantum statistical models for single qubits and we will evaluate the corresponding quantumness parameter $\mathcal{R}$ and we will compare it with the renormalized difference between Holevo and SLD-bound $\Delta C({\bm{\lambda}},{\mathbf W})$. We will look in particular for its maximum value, obtained by varying the weight matrix ${\mathbf W}$, i.e.
\begin{align}
\Delta C_{\sf max} = \max_{{\mathbf W} >0} \Delta C({\bm{\lambda}},{\mathbf W}) \,.
\end{align}
It is important to remark that also this quantity is invariant under reparametrization, as considering a different weight matrix simply corresponds to consider a new set of parameters to be estimated.

We will start by considering the full-tomography case for both pure and mixed qubit states, corresponding respectively to $n=2$ and $n=3$ parameter estimation problems. Then we will approach several different two-parameter models describing the evolution of qubits into noisy channels, such as phase-diffusion and amplitude damping.
\subsection{Pure state model}
Given a generic two-parameter pure state model $|\psi_{\bm{\lambda}}\rangle$, with $\bm{\lambda} = (\lambda_1,\lambda_2)$, the SLD-QFI matrix and the Uhlman curvature matrix can be easily evaluated via the following equations 
\begin{align}
{\bf Q}(\bm{\lambda})=4\left(
\begin{array}{cc}
\alpha+a^2&{\rm Re}(c)+ab\\
{\rm Re}(c)+ab&\beta+b^2
\end{array}
\right)\,, \label{eq:Qpure} \\
{\bf D}(\bm{\lambda}) =4 \left(
\begin{array}{cc}
0& {\rm Im}(c)\\
-{\rm Im}(c)&0
\end{array}
\right)\,, \label{eq:Dpure}
\end{align}
where $a\equiv\langle\partial_{\bm{\lambda}_{1}}\psi_{\bm{\lambda}}|\psi_{\bm{\lambda}}\rangle\,\,$,
$b\equiv\langle\partial_{\bm{\lambda}_{2}}\psi_{\bm{\lambda}}|\psi_{\bm{\lambda}}\rangle\,\,$, $c\equiv\langle\partial_{\bm{\lambda}_{1}}\psi_{\bm{\lambda}}|\partial_{\bm{\lambda}_{2}}\psi_{\bm{\lambda}}\rangle\,\,$,
$\alpha\equiv\langle\partial_{\bm{\lambda}_{1}}\psi_{\bm{\lambda}}|\partial_{\bm{\lambda}_{1}}\psi_{\bm{\lambda}}\rangle\,\,$ and
$\beta\equiv\langle\partial_{\bm{\lambda}_{2}}\psi_{\bm{\lambda}}|\partial_{\bm{\lambda}_{2}}\psi_{\bm{\lambda}}\rangle$.
It follows that
\begin{equation}
\mathcal{R}=\frac{| {\rm Im}(c)|}{\sqrt{(ab+{\rm Re}(c))^2 -4(a^2+\alpha)(b^2+\beta)}} \,.
\end{equation}
\\

\noindent
Here we are interested in a generic pure qubit state
\begin{align}
|\psi_{\bm{\lambda}} \rangle = \cos\frac\theta 2 |0\rangle + e^{i\phi} \sin\frac\theta 2 |1\rangle\,.
\end{align}
and thus in studying the estimation properties of the parameters $\bm{\lambda} = (\theta ,\phi)$, corresponding to the full tomography of the state. By exploiting the Eqs. (\ref{eq:Qpure}) and (\ref{eq:Dpure}), one can easily evaluate the SLD matrices, obtaining
\begin{align}
{\mathbf Q}(\bm{\lambda})&=\left(
\begin{array}{cc}
1&0\\
0&\sin^2\theta
\end{array}
\right)\,,\\ 
{\mathbf D}(\bm{\lambda}) &= \left(
\begin{array}{cc}
0& -\sin\theta \\
\sin\theta&0
\end{array}
\right)\,.
\end{align}
Consequently one obtains $\mathcal{R}=\sqrt{\det {\bf Q} / \det{\bf D}} = 1$, that is, according to this measure, the parameters are maximally incompatible for any possible values of $\theta$ and $\phi$ (notice that we are not considering the case where the SLD-QFI matrix is singular, $\theta=\{0,\pi\}$). 

We are now interested in evaluating the SLD- and the Holevo bound; in particular we will restrict to diagonal weight matrices that can be generically written as ${\mathbf W} = {\rm diag}(1,w)$,  with $w>0$. 
It has been demonstrated that any pure state qubit model is D-invariant \cite{Suzuki2016a}. While in general the RLD operators for a pure state model are not well defined, it was shown in \cite{Fujiwara1999} that the formula in Eq. (\ref{eq:HolevoDInvariant}) is still valid also in the limit of pure states. Consequently, one can evaluate both the SLD- and the Holevo-bound as
\begin{align}
C^{\sf S}({\bm{\lambda}},{\mathbf W}) &= \text{Tr}[{\mathbf W} {\mathbf Q}^{-1}] = 1 + \frac{w}{\sin^2 \theta} \,, \\
C^{\sf H}({\bm{\lambda}},{\mathbf W}) &= \text{Tr}[{\mathbf W} {\mathbf Q}^{-1}] +  \Vert \sqrt{\mathbf{W}}  \, \mathbf{Q}^{-1} \, \mathbf{D} \, \mathbf{Q}^{-1}\, \sqrt{\mathbf{W}} \Vert_1 = \left( 1 + \frac{\sqrt{w}}{\sin\theta} \right)^2 \,,
\end{align}
leading to
\begin{align}
\Delta C({\bm{\lambda}},{\mathbf W}) = \frac{2 \sqrt{w} \sin\theta}{w + \sin^2\theta} \,.
\end{align}
Remarkably one observes that by choosing $w=w^{\sf (max)} = \sin^2\theta$ one has $\Delta C_{\sf max} = \Delta C({\bm{\lambda}},{\mathbf W}) = 1$, that is for any value of the parameters $\bm{\lambda}$, it is possible to find a diagonal weight matrix such that the renormalized difference between Holevo and SLD-bound is equal to the quantumness $\mathcal{R}$ (and in this case to its maximum value 1). Remarkably the optimal weight matrix is equal to the SLD-QFI matrix, $\mathbf{W}=\mathbf{Q}(\bm{\lambda})$, that is when the parameters are weighted according to the Bures metric defined by the statistical model \cite{Braunstein1994}.
We have thus evidence that for this model the bound $\Delta C({\bm{\lambda}},{\mathbf W})  \leq\mathcal{R}$ is in fact tight, and thus the quantity $\mathcal{R}$ is a good measure of incompatibility of the two parameters $\theta$ and $\phi$. On the other hand we also observe how, at fixed weight matrix ${\bf W}$, the difference between Holevo- and SLD-bound may be much smaller than the one predicted by the quantumness $\mathcal{R}$. \\

We finally remark that any other two-parameter estimation problem involving pure qubit states, will be characterized by the same figures of merit $\mathcal{R}$ and $\Delta C _{\sf max}$. We indeed proved via the property {\bf [P5]} that the quantumness parameter $\mathcal{R}$  is invariant under reparametrization. As a consequence the result above holds also for the (unitary) quantum statistical model corresponding to pure states 
\begin{align}
|\psi_{\overline{\bm{\lambda}}} \rangle = e^{i \lambda_x \sigma_1 + i \lambda_z \sigma_3} |\psi_0\rangle \,,\,\,\, 
\overline{\bm{\lambda}} = ( \lambda_x,\lambda_z ) \,,
\end{align}
that is corresponding to the estimation of the two {\em phases} $\lambda_x$ and $\lambda_z$ due to the unitary evolution $U =e^{i \lambda_x \sigma_1 + i \lambda_z \sigma_3} $ applied on a generic initial state $|\psi_0\rangle$. 
\subsection{Full tomography of a qubit mixed state}
We now consider the quantum statistical model corresponding to a generic mixed qubit state
\begin{align}
\varrho_{\bm{\lambda}} = \frac{1}{2} \left( \mathbbm{1} + \sum_{j=1}^3 \gamma_j \sigma_j \right) \,, 
\end{align}
where the matrices $\sigma_j$ denote Pauli matrices and 
\begin{align}
\gamma_1 = r \sin\theta \cos\phi \,, \,\,\,\,\,
\gamma_2 = r \sin\theta \sin\phi \,, \,\,\,\,\,
\gamma_3 = r \cos\theta  \,,
\end{align}
and we thus consider the set of parameters $\bm{\lambda} = ( r ,\theta, \phi)$ characterizing the vector in the Bloch sphere corresponding to the state $\varrho_{\bm{\lambda}}$. The SLD operators can be easily evaluated by solving the corresponding Lyapunov equations, yielding the matrices
\begin{align}
{\mathbf Q}(\bm{\lambda})&=\left(
\begin{array}{ccc}
1/(1-r^2)&0 & 0\\
0& r^2 & 0 \\
0 & & r^2 \sin\theta^2 
\end{array}
\right)\,,\\ 
{\mathbf D}(\bm{\lambda}) &= \left(
\begin{array}{ccc}
0 & 0 & 0 \\
0 & 0 &  r^3 \sin\theta \\
0 & - r^3 \sin\theta&0 
\end{array}
\right)\,.
\end{align}
We will investigate the regime where the SLD-QFI matrix is not singular, that is avoiding the cases of maximally mixed ($r=0$) or maximally pure ($r=1$) states, and for values of the azimuthal angle $\theta=\{0,\pi\}$.
By evaluating the quantumness parameter via its definition (\ref{eq:Rdefinition}) one obtains $\mathcal{R}=r$, that is the quantumness is in fact equal to the length of the Bloch vector characterizing the qubit.

As in the previous case, we now focus on diagonal weight matrices ${\mathbf W} = {\rm diag}(1,w_\theta,w_\phi)$ that can be parametrized in terms of two positive real numbers $w_\theta$ and $w_\phi$. Also in this case one proves that the model is D-invariant and as a consequence one can evaluate both the SLD- and the Holevo bound 
\begin{align}
C^{\sf S}({\bm{\lambda}},{\mathbf W}) &= \frac{(r^2 - r^4 + w_\theta) \sin^2\theta + w_\phi}{r^2 \sin^2 \theta} \,, \\
C^{\sf H}({\bm{\lambda}},{\mathbf W}) &= \frac{(r^2 - r^4 + w_\theta) \sin^2\theta +2 r \sqrt{w_\theta w_\phi} \sin\theta + w_\phi}{r^2 \sin^2 \theta} \,,
\end{align}
corresponding to
\begin{align}
\Delta C({\bm{\lambda}},{\mathbf W}) = \frac{2 r \sqrt{w_\theta w_\phi} \sin\theta}{(r^2 - r^4 + w_\theta) \sin^2\theta + w_\phi} \,.
\end{align}
We observe 
One can then check that it is not possible to find a couple of real positive weights $(w_\theta, w_\phi)$ such that 
$\Delta C({\bm{\lambda}},{\mathbf W}) = r$, unless we have $r=1$, that is for pure states, as already shown in the previous section. In particular, we also notice that in this case, by choosing as the weight matrix the SLD-QFI matrix, $\mathbf{W}=\mathbf{Q}(\bm{\lambda})$, one obtains $\Delta C({\bm{\lambda}},{\mathbf W})=2r/3 = 2\mathcal{R}/3$, and thus weighting the parameters according to the Bures metric is not the optimal choice. 
However we can check that for generic $r$, by setting $w_\theta$ equal to the real part of the solution of the equation $\Delta C({\bm{\lambda}},{\mathbf W}) = r$, that is
\begin{align}
w_\theta^{\sf (max)} = \frac{w_\phi + (r^4 - r^2) \sin^2 \theta}{\sin^2 \theta} \,,
\end{align}
one obtains
\begin{align}
\Delta C({\bm{\lambda}},{\mathbf W}) = \frac{r \sqrt{w^2_\phi + w_\phi  (r^4 - r^2) \sin^2 \theta}}{w_\phi} \,,
\end{align}
that, in the limit of $w_\phi\rightarrow \infty$, gives $\Delta C({\bm{\lambda}},{\mathbf W}) \rightarrow r$. However, as also suggested by some numerics, it is not necessary to consider an {\em infinite weight} $w_\phi$, as it is easy to find diagonal weight matrices ${\mathbf W}$ such that the corresponding renormalized difference $\Delta C({\bm{\lambda}},{\mathbf W})$ is $\epsilon$-close to the quantumness parameter $\mathcal{R}=r$. In general we can conclude that also in this case the bound for $\Delta C({\bm{\lambda}},{\mathbf W})$ is {\em almost} tight and thus the parameter $\mathcal{R}$ is in fact a good measure of quantumness for the statistical model.
\subsection{Simultaneous estimation of frequency and dephasing rate}
We now consider the evolution of a qubit system corresponding to a simultaneous rotation around the $z$-axis with frequency $\omega$ and dephasing with rate $\gamma$, ruled by the following Lindblad master equation
\begin{equation}
\dot{\varrho}=-i\frac{\omega}{2}  [\sigma_3,\varrho]+\frac{\gamma}{2}\mathcal{D}[\sigma_3]\varrho \,, \label{eq:medeph}
\end{equation}
where we have defined the superoperator 
\begin{equation}
\mathcal{D}[A]\varrho = A \varrho A^\dag - \frac{1}{2}\left( A^\dag A \varrho + \varrho A^\dag A \right) \,. 
\end{equation}
Given an generic initial pure state $|\psi_0\rangle=\cos({\theta/2})|0\rangle+e^{i\phi}\sin({\theta/2})|1\rangle$ the evolved density matrix can be analytically evaluated, and we are going to consider it as our quantum statistical model
\begin{equation}
\varrho_{\bm{\lambda}}=\left(
\begin{array}{cc}
\cos^2({\theta/2})&\frac{1}{2}e^{-(\gamma-i\omega)t-i\phi}\sin{\theta}\\
\frac{1}{2}e^{-(\gamma+i\omega)t+i\phi}\sin{\theta}&\sin^2({\theta/2}) 
\end{array}
\right), \,\,\,\,\, {\bm{\lambda}} = (\omega, \gamma ) \,.
\end{equation}
As it clear from the formula above, the evolution time $t$ is just a multiplicative factor for both the parameters we  want to estimate, that is the frequency $\omega$ and the dephasing rate $\gamma$, and thus will not play any fundamental role in the evaluation of our figures of merit. The SLD- and the RLD-operators can be evaluated without difficulties by solving the corresponding defining equations (\ref{eq:SLD}) and (\ref{eq:RLD}). The corresponding QFI matrices and the Ulhman curvature matrix then reads 
\begin{align}
{\mathbf Q}(\bm{\lambda}) &=\left(
\begin{array}{cc}
\frac{4 t^2 \sin^2 \theta}{e^{2 \gamma t} - 1}
&0\\
0&4 e^{-2\gamma t} t^2 \sin^2\theta
\end{array}
\right)\,,\\ 
{\mathbf D}(\bm{\lambda}) &= \left(
\begin{array}{cc}
0& 
4 e^{-2\gamma t} t^2 \,\cos\theta \sin^2\theta  \\
-4 e^{-2\gamma t} t^2 \,\cos\theta \sin^2\theta 
&0
\end{array}
\right)\,, \\
{\mathbf J}(\bm{\lambda}) &= \left(
\begin{array}{cc}
\frac{ 4 t^2 }{e^{2 \gamma t} -1} & 
\frac{4 i t^2 \cos\theta}{e^{2 \gamma t} -1}  \\
-\frac{4 i t^2 \cos\theta}{e^{2 \gamma t} -1}  &
\frac{ 4 t^2 }{e^{2 \gamma t} -1}
\end{array}
\right)\,.
\end{align}
By exploiting Eq. (\ref{eq:R2parameter}), we obtain quantumness parameter $\mathcal{R}$ 
\begin{equation}
\mathcal{R}= \lvert \cos\theta \rvert \sqrt{1-e^{-2 \gamma t}} \,,
\end{equation}
showing how the maximum incompatibility at fixed $\gamma$ is obtained in the limit of $\theta\to 0$ and $\theta \to \pi$, that is when the model is not well defined, as the SLD-QFI is singular. In this limit in fact the initial state is an eigenstate of $\sigma_3$ and thus the state remains unchanged during the evolution due to the master equation  (\ref{eq:medeph}) without acquiring any information on the parameters. As regards the behaviour of $\mathcal{R}$ as a function of the dephasing parameter $\gamma$, it is easy to check that $\mathcal{R}$ monotonically increases with $\gamma$, and thus by decreasing the purity of the corresponding quantum statistical model $\varrho_{\bm{\lambda}}$; this observed behavior is opposite to what we have discussed previously for state tomography, where we in fact found that the quantumness $\mathcal{R}$ coincided with the purity of the qubit state $\varrho_{\bm{\lambda}}$. 

\begin{figure}[t!]
	\centering
		\includegraphics[width=\textwidth]{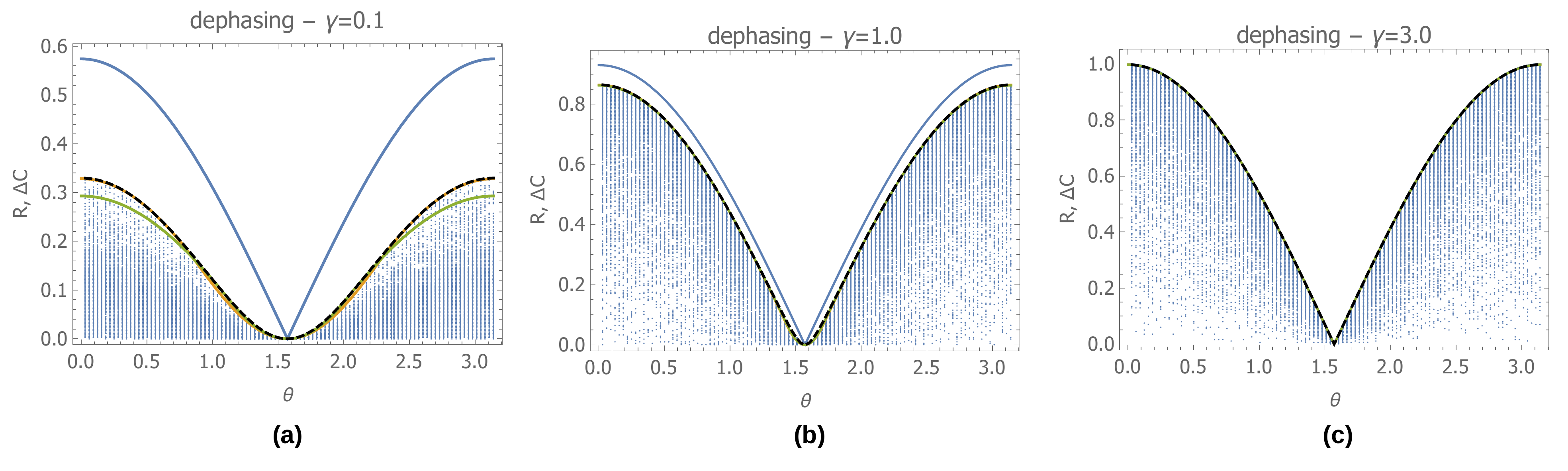}
	\caption{Plots of the quantumness parameter $\mathcal{R}$ (blue solid line) and of the renormalized difference $\Delta C(\bm{\lambda}, {\mathbf W})$ for simultaneous estimation of frequency and dephasing rate, as a function of the initial state parameter $\theta$. The black dashed line correspond to $\Delta C(\bm{\lambda}, {\mathbf W})$ with an optimized diagonal weight matrix ${\mathbf W}$ for every value of $\theta$. Yellow and green correspond to $\Delta C(\bm{\lambda}, {\mathbf W})$ with diagonal weight matrix optimized respectively for $\theta=\{\pi/3, 2\pi/3\}$ (in some of the plots these curves are not visible as they are perfectly superimposed by the dashed-black line corresponding to the optimized $\Delta C$). Blue points corresponds to $\Delta C(\bm{\lambda}, {\mathbf W})$ evaluated for random generic weight matrices.\\
		The three plots correspond to different values of the dephasing rate: (a) $\gamma t = 0.1$; (b) $\gamma t=1.0$; (c) $\gamma t=3.0$.}%
	\label{f:dephasing}%
\end{figure}

For this quantum statistical model the Holevo bound has to be evaluated via Eq. (\ref{eq:Holevoqubit}), as the model is neither asymptotically classical, nor D-invariant. In particular, as in the previous examples, we started by considering diagonal weight matrices ${\mathbf W}= {\rm diag}(1, w)$. The optimal $w=w^{\sf (max)}$ maximizing $\Delta C(\bm{\lambda}, {\mathbf W})$ depends on the initial state of the qubit $|\psi_0\rangle$, and in particular on its angle $\theta$. Remarkably we have strong numerical evidence that in the limit $\theta \to 0$ or $\theta \to \pi$, we have $w^{\sf max} = \Delta C(\bm{\lambda},{\mathbf W})^2$, that is exactly equal to the square of the corresponding renormalized difference between Holevo and SLD-bound.
We have also generated random generic (non-diagonal) weight matrix and, as one can see in Fig. \ref{f:dephasing}, we obtain that optimizing over diagonal ${\mathbf W}$ is enough to obtain $\Delta C_{\sf max}$, that is in general one obtains values of $\Delta C$ smaller than the one optimized on diagonal-weight matrices (we also notice that $\mathbf{W}=\mathbf{Q}({\bm{\lambda}})$ is not optimal at fixed $\gamma$). Moreover we also observe that $\Delta C_{\sf max}$ and $\mathcal{R}$ share the same qualitative behaviour as a function of the parameters $\theta$ and $\gamma$. However, we also observe that for small values of the dephasing rate $\gamma$, $\Delta C$ is strictly smaller than $\mathcal{R}$, that is the quantumness parameter seems to overestimate the incompatibility of frequency and dephasing. Only by increasing $\gamma$ we find that the gap between these two quantities goes to zero for all values of $\theta$ (in particular we find that this is already the case for $\gamma=1.5$). In the limit of $\gamma \to \infty$, the optimal weight matrix becomes the identity ${\mathbf W} =\mathbbm{1}$, a choice that is equivalent to weighting the parameters according to the Bures metric (in this limit the ratio between the two diagonal non-zero elements of the SLD-QFI matrix $[\mathbf{Q}({\bm{\lambda}})
]_{2,2}/[\mathbf{Q}({\bm{\lambda}})]
_{1,1}$ goes indeed to one).
\subsection{Simultaneous estimation of frequency and amplitude damping rate}
In this last example we consider a different noisy evolution, ruled by the Lindblad master equation
\begin{equation}
\dot{\varrho}=-i\frac{\omega}{2}  [\sigma_3,\varrho]+{\gamma}\mathcal{D}[\sigma_-] \varrho\,, \,\,\,\,\,\,\textrm{where}\,\, \sigma_{\tiny -} = \frac{\sigma_1 - i \sigma_2}{2}\,,
\end{equation}
corresponding to simultaneous rotation around the $z$-axis with frequency $\omega$ and amplitude damping with rate $\gamma$. Also in this case the equation can be analytically solved, and for a generic initial state $|\psi_0\rangle=\cos({\theta/2})|0\rangle+e^{i\phi}\sin({\theta/2})|1\rangle$, we obtain our quantum statistical model
\begin{equation}
\varrho_{\bm{\lambda}}=\left(
\begin{array}{cc}
1-e^{\gamma t}\sin^2(\theta/2)& \frac{1}{2}e^{-(\gamma/2+i\omega)t-i\phi}\sin{\theta}\\
\frac{1}{2}e^{-(\gamma/2-i\omega)t+i\phi}\sin{\theta}&e^{\gamma t}\sin^2(\theta/2)
\end{array}
\right), \,\,\,\,\, {\bm{\lambda}} = ( \omega, \gamma ) \,.
\end{equation}
As in the previous example the evolution time $t$ is just a multiplicative factor for both parameters and thus is not going to play any role in our results.

The equations for the SLD- and the RLD-operators can be easily solved, and consequently we have been able to evaluate analytically the following matrices
\begin{align}
{\mathbf Q}(\bm{\lambda}) &=\left(
\begin{array}{cc}
\frac{e^{-\gamma t} t^2 \sin^2(\frac{\theta}{2}) \left(1 + \cos\theta - 2 e^{\gamma t} \right)}{2 ( 1-e^{\gamma t} )} 
&0\\
0&4 e^{-\gamma t} t^2 \sin^2\theta
\end{array}
\right)\,,\\ 
{\mathbf D}(\bm{\lambda}) &= \left(
\begin{array}{cc}
0& 
e^{-2\gamma t} t^2 \,\sin^2\theta ( \cos\theta - 1 - e^{\gamma t} ) \\
-e^{-2\gamma t} t^2 \, \sin^2\theta ( \cos\theta - 1 - e^{\gamma t} )
&0
\end{array}
\right)\,, \\
{\mathbf J}(\bm{\lambda}) &= \left(
\begin{array}{cc}
\frac{ t^2 (e^{\gamma t} +  4 + (e^{\gamma t} - 4) \cos\theta )}{8 \sin^2(\frac{\theta}{2}) ( e^{\gamma t} - 1)}  & 
\frac{ i t^2 \cos^2 (\frac{\theta}{2}) ( \cos\theta - 1 - e^{\gamma t})}{(e^{\gamma t}-1)\sin^2 (\frac{\theta}{2}) } \\
-\frac{ i t^2 \cos^2 (\frac{\theta}{2}) ( \cos\theta - 1 - e^{\gamma t})}{(e^{\gamma t}-1)\sin^2 (\frac{\theta}{2}) }
&
\frac{4 e^{\gamma t} t^2 \cos^2 (\frac{\theta}{2} )}{e^{\gamma t} - 1} 
\end{array}
\right)\,.
\end{align}
The quantumness parameter $\mathcal{R}$ can be straightforwardly evaluated by exploiting Eq. (\ref{eq:R2parameter}), obtaining
\begin{equation}
\mathcal{R}=e^{-\gamma t}\cos\frac{\theta}{2} (1+e^{\gamma t}-\cos{2\theta})\sqrt{\frac{2(1-e^{\gamma t})}{1-2e^{\gamma t}+\cos{\theta}}} \,.
\end{equation}
As it is clear from the equations, all these quantities, and in particular the parameter $\mathcal{R}$ depend only on the initial angle $\theta$ and on the amplitude damping $\gamma t$. In particular, at fixed $\gamma t$, one can show that $\mathcal{R}$ has a monotonous decreasing behaviour with $\theta$, from its maximum value $\mathcal{R}=1$ for $\theta \to 0$, to its minimum value $\mathcal{R}=0$ for $\theta \to \pi$. It is important to notice that these extremal cases correspond to the values of $\theta$ that make the SLD-QFI singular (for $\theta \to 0$ all the elements of the SLD-QFI matrix actually become identical to zero). As regards the behaviour as a function of $\gamma$, we observe that $\mathcal{R}$ is monotonically increasing for $\gamma t \in [0,\ln 2]$, and then monotonically decreasing in the interval $\gamma t \in (\ln 2, \infty)$. As in the previous example, we find that this behaviour is opposite to the behaviour of the purity of the quantum state $\varrho_{\bm{\lambda}}$, which is indeed decreasing for $\gamma t\in [0,\ln 2]$ and then increasing for larger values of $\gamma$.

\begin{figure}[h!]
	\centering
		\includegraphics[width=\textwidth]{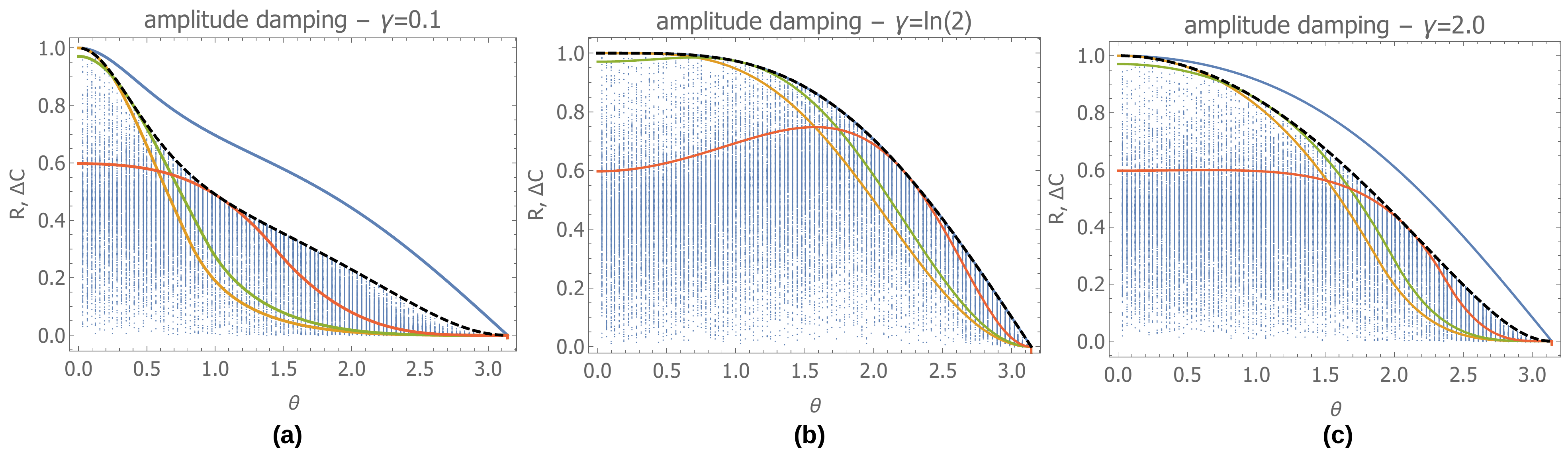}
	\caption{Plots of the quantumness parameter $\mathcal{R}$ (blue solid line) and of the renormalized difference $\Delta C(\bm{\lambda}, {\mathbf W})$ for simultaneous estimation of frequency and amplitude damping rage, as a function of the initial state parameter $\theta$. The black dashed line corresponds to $\Delta C(\bm{\lambda}, {\mathbf W})$ with an optimized diagonal weight matrix ${\mathbf W}$ for every value of $\theta$. Yellow, green and red lines correspond to $\Delta C(\bm{\lambda}, {\mathbf W})$ with diagonal weight matrix optimized respectively for $\theta=\{0, \pi/3, 2\pi/3\}$. Blue points corresponds to $\Delta C(\bm{\lambda}, {\mathbf W})$ evaluated for random generic weight matrices.\\
		The three plots correspond to different values of the amplitude damping rate: (a), $\gamma t = 0.1$; (b), $\gamma t=\ln 2$; (c), $\gamma t=2.0$.}%
	\label{f:amplitudedamping}%
\end{figure}

This quantum statistical model is neither asymptotically classical, nor D-invariant. Consequently, the Holevo-bound and the renormalized difference $\Delta C(\bm{\lambda}, {\mathbf W})$ have been evaluated numerically by exploiting Eq. (\ref{eq:Holevoqubit}). We started again our investigation by considering a diagonal weight matrix ${\mathbf W} = {\rm diag}(1,w)$, and we have found that, also by optimizing over the free parameter $w$ for fixed initial state $|\psi_0\rangle$, the renormalized difference $\Delta C(\bm{\lambda}, {\mathbf W})$ is in general smaller than the quantumness $\mathcal{R}$. As before, in order to assess the generality of this result we have generated numerically thousands of non-diagonal random weight matrices. As one can observe from Fig. \ref{f:amplitudedamping}, we have observed that in general the maximum value of $\Delta C$ obtained via diagonal weight matrices is in fact an upper bound for generic weight matrices. We thus assume that by optimizing over diagonal matrices one obtains $\Delta C_{\sf max}$. Observing the figure we have evidence that also in this case the qualitative behaviour of $\Delta C_{\sf max}$ and $\mathcal{R}$ as a function of the different parameters is the same; however it is possible to observe a non-zero gap between $\mathcal{R}$ and $\Delta C$, and thus the parameter $\mathcal{R}$ in general overestimates the degree of incompatibility of the two parameters. In particular, we observe that the gap is always closed in the limit of $\theta \to 0$,  that is when $\mathcal{R}=1$, and the optimal weight matrix parameter takes the value $w\approx 16$. We observe that the gap is also closed for all values of $\theta$ by taking $\gamma t=\ln 2$, where we obtain
\begin{align}
\mathcal{R} = \Delta C_{\sf max} = \cos (\theta/2) \sqrt{\frac{3-\cos\theta}{2}} \,.
\end{align}
In this case the optimal (diagonal) weight parameter reads
\begin{align}
w^{\sf (max)} = \frac{[\mathbf{Q}({\bm{\lambda}})]_{2,2}}{[\mathbf{Q}({\bm{\lambda}})]_{1,1}} = \frac{16 ( 1 + \cos\theta)}{3- \cos\theta} \,,
\end{align}
that is the optimality is obtained by weighting the parameters according to the corresponding Bures metric.  We have thus found another instance where the bound is tight, and the optimal weight matrix is equal (or equivalently proportional) to the SLD-QFI matrix  $\mathbf{Q}({\bm{\lambda}})$ (we also remark that for values of $\gamma t \neq \ln 2$, that is whenever the bound is not tight, the optimal weight matrix differs from $\mathbf{Q}({\bm{\lambda}})$).
\subsection{Asymptotically classical models}
Here we will list a couple of examples of Lindblad master equations for qubits involving two parameters, whose solutions correspond to quantum statistical models $\varrho_{\bm{\lambda}}$ that are asymptotically classical, that is whose quantumness parameter $\mathcal{R}$ is equal to zero. 
\begin{itemize}
	\item {\em Simultaneous estimation of frequency and depolarizing channel rate}, corresponding to the master equation
	\begin{align}
	\dot{\varrho}=-i \frac{\omega}{2}[\sigma_3,\varrho]+\frac{\gamma}{2}\left(\frac{1}{3}\sum_{i=1}^{3}{\sigma_i\,\varrho\,\sigma_i-\varrho}\right),  \,\,\,\,\, {\bm{\lambda}} = ( \omega, \gamma )
	\end{align}
	\item {\em Simultanoues estimation of amplitude damping and dephasing rates}, corresponding to the master equation
	\begin{align}
	\dot{\varrho}=\gamma_{\sf ad}\mathcal{D}[\sigma_{-}]\varrho+\frac{\gamma_{\sf deph}}{2} \mathcal{D}[\sigma_{3}]\varrho,  \,\,\,\,\, {\bm{\lambda}} = ( \gamma_{\sf ad}, \gamma_{\sf deph} )
	\end{align}
\end{itemize}

\section{Discussion and conclusions}
In this paper, we have studied in detail the quantumness of multiparameter 
quantum statistical models for qubit systems, defined as the incompatibility 
of the parameters to be jointly estimated. 
In particular, we have evaluated  the renormalized 
difference $\Delta C_{\sf max}$ between the Holevo- and the SLD-bound optimized over all the possible weight matrices,  
and its upper bound given by the quantumness measure $\mathcal{R}$. 
Our results confirm that $\mathcal{R}$ is a useful and practical tool 
to characterize the properties of the quantum statistical model: (i) we have shown some examples where in fact $\Delta C_{\sf max} = \mathcal{R}$, and remarkably we have found that in these cases the weight matrix maximizing $\Delta C$ always corresponds to the Bures metric induced by the quantum statistical model; (ii)  we have observed that in general the two quantities, $\mathcal{R}$ and $\Delta C_{\sf max}$, have the same qualitative behaviour. In particular they show a peculiar counterintuitive dependence on the purity of the quantum states $\varrho_{\bm{\lambda}}$: in the quantum state tomography scenario, both $\mathcal{R}$ and $\Delta C_{\sf max}$ are monotonically increasing with the purity, while in the two noisy models induced by the Markovian master equation, we observe the opposite behaviour, and larger values of $\mathcal{R}$ (or $\Delta C_{\sf max}$) are obtained for low purity states. \\
However our results give also clear evidence that the bound is in general not always tight, i.e. it is possible to find several examples where $\mathcal{R}$ overestimates the actual degree of incompatibility of the parameters, and thus the evaluation of the Holevo-bound is needed to properly assess and quantify this property of the quantum statistical model.

We believe that our work together with other complementary approaches, such as the one pursued in \cite{Kull2020} where trade-off surfaces are derived via the SLD- and the Holevo-bound, will help in shedding new light on the relationship between quantum uncertainty relations, incompatibility and multi-parameter quantum metrology.

\begin{acknowledgments}
MGG acknowledges support by MIUR (Rita Levi-Montalcini fellowship). MGAP~is a member of INdAM-GNFM. The authors acknowledge several useful discussions with Francesco Albarelli.
\end{acknowledgments}

\bibliography{persbiblio}
\clearpage


\clearpage
\setcounter{figure}{0}
\text{Re}newcommand\thefigure{S\arabic{figure}}    
\end{document}